\documentclass[aoas,MSNbibl,nameyear,seceqn,rotating,dvips]{arximspdf}
\usepackage{dcolumn}
\usepackage{graphicx}

\doi{10.1214/12-AOAS618} 
\volume{7}
\issue{2}
\pubyear{2013}
\firstpage{613}
\lastpage{639}

\makeatletter

\newcolumntype{d}[1]{D{.}{.}{#1}}

\makeatother

\begin{document}
\begin{frontmatter}

\title{Model trees with topic model preprocessing:
An~approach for data journalism illustrated with the WikiLeaks
Afghanistan war logs}
\runtitle{Model trees with topic model preprocessing}

\begin{aug}
\author[A]{\fnms{Thomas}~\snm{Rusch}\corref{}\ead[label=e1]{thomas.rusch@wu.ac.at}},
\author[B]{\fnms{Paul}~\snm{Hofmarcher}\ead[label=e2]{paul.hofmarcher@jku.at}},
\author[C]{\fnms{Reinhold}~\snm{Hatzinger}\thanksref{t1}\ead[label=e3]{reinhold.hatzinger@wu.ac.at}}
\and
\author[D]{\fnms{Kurt}~\snm{Hornik}\ead[label=e4]{kurt.hornik@wu.ac.at}}
\runauthor{Rusch, Hofmarcher, Hatzinger and Hornik}
\affiliation{WU Vienna University of Economics and Business,
Johannes Kepler University Linz, WU Vienna University of Economics and
Business and WU Vienna University of Economics and Business}
\address[A]{T. Rusch\\
Center for Empirical Research Methods\\
WU (Vienna University of Economics\\
\quad and Business)\\
Augasse 2-6\\
1090 Vienna\\
Austria\\
\printead{e1}}
\address[B]{P. Hofmarcher\\
Institute for Applied Statistics\hspace*{38pt}\\
Johannes Kepler University\\
Altenbergerstrasse 69 \\
4040 Linz\\
Austria\\
\printead{e2}}
\address[C]{R. Hatzinger\\
Center for Empirical Research Methods\\
WU (Vienna University of Economics\\
\quad and Business)\\
Augasse 2-6\\
1090 Vienna\\
Austria\\
and\\
Institute for Statistics and Mathematics\\
Department of Finance, Accounting\\
\quad and Statistics\\
WU (Vienna University of Economics\\
\quad and Business)\\
Augasse 2-6\\
1090 Vienna\\
Austria\\
\printead{e3}}
\address[D]{K. Hornik\\
Institute for Statistics and Mathematics\\
Department of Finance, Accounting\\
\quad and Statistics\\
WU (Vienna University of Economics\\
\quad and Business)\\
Augasse 2-6\\
1090 Vienna\\
Austria\\
\printead{e4}} 
\end{aug}

\thankstext{t1}{Professor Hatzinger passed away in July 2012 while the
manuscript was revised.}

\received{\smonth{12} \syear{2011}}
\revised{\smonth{9} \syear{2012}}

%
\begin{abstract}
The WikiLeaks Afghanistan war logs contain nearly $77\mbox{,}000$ reports of
incidents in the US-led Afghanistan war, covering the period from
January 2004 to December 2009. The recent growth of data on complex
social systems and the potential to derive stories from them has
shifted the focus of journalistic and scientific attention increasingly
toward data-driven journalism and computational social science. In this
paper we advocate the usage of modern statistical methods for problems
of data journalism and beyond, which may help journalistic and
scientific work and lead to additional insight. Using the WikiLeaks
Afghanistan war logs for illustration, we present an approach that
builds intelligible statistical models for interpretable segments in
the data, in this case to explore the fatality rates associated with
different circumstances in the Afghanistan war. Our approach combines
preprocessing by Latent Dirichlet Allocation (LDA) with model trees.
LDA is used to process the natural language information contained in
each report summary by estimating latent topics and assigning each
report to one of them. Together with other variables these topic
assignments serve as splitting variables for finding segments in the
data to which local statistical models for the reported number of
fatalities are fitted. Segmentation and fitting is carried out with
recursive partitioning of negative binomial distributions. We identify
segments with different fatality rates that correspond to a small
number of topics and other variables as well as their interactions.
Furthermore, we carve out the similarities\vadjust{\goodbreak} between segments and connect
them to stories that have been covered in the media. This gives an
unprecedented description of the war in Afghanistan and serves as an
example of how data journalism, computational social science and other
areas with interest in database data can benefit from modern
statistical techniques.
\end{abstract}

%

\end{frontmatter}

\section{Introduction}
Analyses of fatalities in wars and armed conflicts are an eminent
subject of systematic investigation. Most of them have been conducted
in a historical context, often retrospectively estimating the number of
and circumstances under which fatalities of war occurred. There are
literally hundreds of historical investigations into numerous wars;
see, for example, \citet{Garfield1991} for a review of the last 200 years.

Notwithstanding such efforts, contemporary systematic scientific
investigation into the number of fatalities in wars are relatively rare
and more closely tied to the emergence of statistics and epidemiology
as disciplines rather than to the discipline of history. Some of the
first examples we could find were \citet{Marshall1838} or
\citet{night1863}. While these investigations were still firmly
rooted in descriptive statistics, statistical modeling was about to
become imperative as
\citet{Bortkiewicz1898} published his seminal work on the use of the
Poisson distribution for rare events which he motivated by the
analysis of deaths of Prussian soldiers by horse kicks. To our knowledge,
this was the first instance of a parametric and inferential approach
to analyze fatalities of war. Contemporary investigations into the
number and circumstances of casualties of war that made use of
statistical modeling next to descriptive approaches have increased
since then, for example, \citet{Spiegel2000}, \citet{Thomas2001},
\citet{Lakstein2005} or \citet{Holcomb2007}.

In the last decade their number seems to peak\setcounter{footnote}{1}\footnote{According to a
quick survey in the ISI Web of Knowledge citation database, searching
for ``war casualties'' in March 2011 found 1476 records, 840 of which
were published after 2000. 580 of those were published no earlier than
2005.} arguably because data on war fatalities are much easier to come
by. Recent work, for example, for the war in Afghanistan, includes the
studies on child casualties by \citet{Bhutta2002} and on military
fatalities by \citet{Bird2007} or \citet{Bohannon2011}. Other recent
work in this field has been done by \citet{Haushofer2010,Degomme2010,buzzell2007,Burnham2006}.

In July 2010 the availability of data on a specific war became
unprecedented, as whistleblower website WikiLeaks released a massive
amount of military classified war logs from the Afghanistan war into
the public. These documents constitute a ``war diary'' of the military
operation in Afghanistan, containing a detailed description of what
happened in each event for which a report was filed, including counts
of killed and wounded people, local and administrative information,
temporal and spatial information and a short written description of
each particular incident. The documents themselves stem from a database
of the US army and, along the lines of WikiLeaks, they do not generally
cover any top secret operations or European or other operations of the
International Security Assistance Force (ISAF). In total, the war logs
consist of $76\mbox{,}911$ documents and cover the time period between
January $2004$ and December $2009$. They provide an unprecedented view
of the war in Afghanistan with an information abundance that has
previously been unknown and has only been topped by the release of the
Iraq war logs some months later.

Interestingly, the scientific community has been rather hesitant in
approaching the data [but see \citet{OLoughlin2010,drewconway2010a,zammit2012} for notable exceptions]. In journalism
and the media world, however, the impact of the release was very
strong. The German news magazine \textit{Der Spiegel} wrote that the
editors-in-chief of \textit{Der Spiegel}, \textit{The New York Times}
and \textit{The Guardian} were ``unanimous in their belief that there
is a justified public interest in the material'' [\citet{spiegel2010}]
and the war diary was marked as the $21$st century equivalent of the
Pentagon Papers from the 1970s. However, while the Pentagon Papers have
provided an aggregated view on the war in Vietnam, the WikiLeaks war
diary is an account of the daily events in Afghanistan containing
thousands of mosaic tiles describing incidents from the perspective of
the US forces. They were written by different people and are sometimes
accurate and sometimes possibly not. The war logs themselves neither
contain information on strategic decisions nor do they provide a
coherent, general picture of the war. Hence, each media outlet had to
write its own stories based on the material
[see \citet{OLoughlin2010}]. This take on the WikiLeaks Afghanistan
war logs has been praised as data-driven journalism in action
[see \citet{guardian2010}].\looseness=1

To elicit stories out of complex data is a contemporary issue for
journalists and (social) scientists, especially when the amount of data
is large and cannot be processed
easily by humans. This is where data journalism or database
journalism (a~type of journalism which allows stories to enfold from
data) and computational social science [the science that investigates
social phenomena through advanced information processing
technologies, e.g., \citet{cioffi2010}] come into play. Data
journalism and computational social science both use statistical and
computational methods to deal with the problem of processing large and
complex data (often in the form of text documents) and presenting them
in an accessible form. For example, a popular approach is to narrow
down the data by keyword searches with the goal to find a relevant
subset that can be processed by a human reader. Another one is to count
the frequency of words within documents to allow for a broad overview
of the data or to extract additional information that can be used for
telling a story without the need for directly reading or processing all
data points
[see, e.g., \citet{hofmarcher2011,Cohen2011}]. More advanced
approaches may aim at clustering the documents into ``similar'' sets of
documents, for example, via bag of words models
[see \citet{Zhang2010}]. This allows the journalist or scientist to
find the story by reading just a few documents within each cluster.
Another approach might be to derive structure from unstructured data
by, for example, using network analysis
[e.g., \citet{lazer2009}] and similar methods. Often a description or
a visualization is the primary goal of such procedures, but in
principle the analysis is not limited to that.\looseness=1 

Regarding the WikiLeaks Afghanistan war logs, analyses up to the point
of writing this paper have remained mostly on a descriptive level and
if insights from an inferential or modeling approach have been gained,
it was mostly by using a small amount of the information available.
This could be due to the nature and bulk of the data. One of the
peculiarities of the war log and its main challenge is that the data at
hand stem from a database and that the information is captured in both
numeric variables as well as written text. To neglect the written text
in a statistical evaluation of such data sets would often come along
with discarding important, if not crucial, information. Especially in
the WikiLeaks data, nearly all detailed information about the events is
stored as written text. Thus, it is essential for a deep statistical
probing to incorporate that information.

Modern statistical procedures provide tools to handle, analyze and
model such data sets appropriately and therefore allow a more thorough
investigation. In this paper we will make exemplary use of statistical
learning procedures to segment the reports in the war logs and to build
local statistical models for the number of fatalities in each segment.
By combining two modern ideas, topic models and model-based recursive
partitioning, our analysis allows to draw a bigger picture of the war
from the thousands of mosaic tiles. In doing so, we present an approach
that might be particularly suitable for, but not limited to, data
journalism and social science, especially since in the end it provides
palpable segments of data points characterized by a small number of
parameters that directly relate to the question at hand.\looseness=1

The idea of our approach is as follows: each single entry in the
WikiLeaks war logs contains several variables and also a written report
summary containing a short description of what happened in the
particular incident. We are interested in extracting explanatory
information from the reports, some type of meta information that
aggregates reports with similar content. We achieve this by using
Latent Dirichlet Allocation [LDA; \citet{Blei2003}] which clusters
written report summaries into latent topics. In a second step, we then
use the generated topic assignments as variables from which we infer a
segmentation of the reports and locally model the number of fatalities
in each segment. The provided fatality counts function as our target
variable. Since there is a high degree of overdispersion present, we
use a negative binomial distribution [\citet{lawless87}] model in each
segment. This enables to estimate the distribution of deaths per
segment appropriately. To allow for a flexible, nonlinear,
interaction-focused functional relationship between splitting variables
and the local model, we employ the model-based recursive partitioning
framework of \citet{zeileis08}.\looseness=1

The remainder of this paper is organized as follows: Section \ref
{secwardiary} contains a description of the WikiLeaks war logs. The
methodological Section \ref{secmethod} presents the methods used. The
results are described and discussed in Section \ref{secresults}. In
Section \ref{secval1} we provide validation of the results. We finish
with conclusions in Section \ref{secConclusion}. This paper is
accompanied by supplementary material
[Rusch et~al. (\citeyear{supplementa,supplementb})].

\section{The WikiLeaks Afghanistan war logs}
\label{secwardiary} 
\begin{figure}[b]

\includegraphics{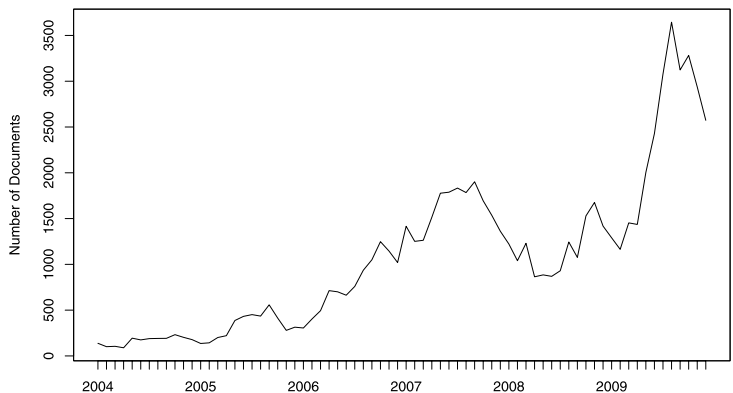}

\caption{Monthly quantity of filed reports.}
\label{figdocuments}
\end{figure}

The release of $76\mbox{,}911$ individual war logs
by \href{http://www.wikileaks.org/}{WikiLeaks.org} provides an
unprecedented possibility to take a look at an ongoing war. The war
logs cover the period from January 2004 to December 2009 and each event
for which a report has been filed corresponds to a single document.
Figure \ref{figdocuments} displays the number of filed reports per
month. While for the first years of the military operation we can find
only a few hundred reports per month, this number increases up to more
than $3500$ per month in mid $2009$.

%

Each report contains $32$ numerical and factor variables. They include
four variables listing the number of ``Civilian,'' ``Enemy,''
``Friend'' and ``Host'' fatalities within each report. The sum of these
fatalities for each report serves as our target variable. Note that
fighters opposing coalition troops are referred to as ``Enemies.'' We
adopt the term ``Anti Coalition Fighters'' (ACF) to denote this
variable. The ``Friends'' column refers to ISAF forces including the
NATO countries and the US military, while ``Host'' stands for local
(Afghan) military and police. We subsume the former under ``coalition
troops'' or ``allied forces'' and the latter under ``Afghan or host
forces.''\looseness=1

\begin{table}
\tablewidth=229pt
\caption{The number of casualties by group}
\label{tabsumm}
\begin{tabular*}{\tablewidth}{@{\extracolsep{\fill}}lrrrrc@{}}
\hline
& \textbf{Allied} & \textbf{Host} & \textbf{Civilian} & \multicolumn{1}{c}{\textbf{ACF}}
& \textbf{Total} \\
\hline
Killed & 1146 & 3796 & 3994 & 15\mbox{,}219 & 24\mbox{,}155 \\
Wounded & 7296 & 8503 & 9044 & 1824 & 26\mbox{,}667 \\
\hline
\end{tabular*}
\end{table}

\begin{figure}[b]

\includegraphics{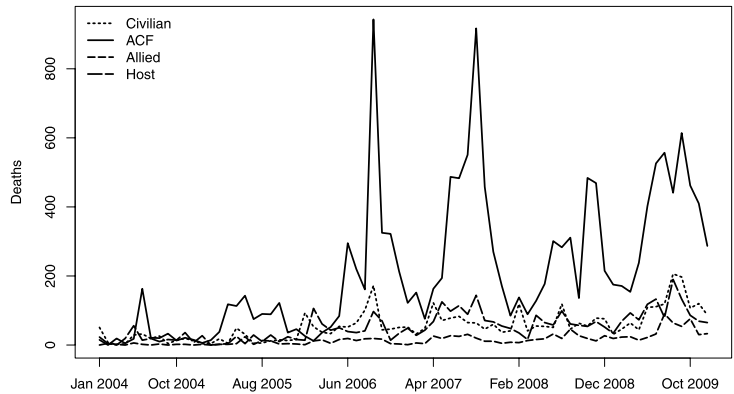}

\caption{Monthly counts of fatalities by group.}
\label{figdeaths}
\end{figure}

Table \ref{tabsumm} provides summary statistics of the reported
casualties and Figure~\ref{figdeaths} displays a plot of the number
of fatalities over time for each group during the observation period.
In total we find $24\mbox{,}155$ fatalities in the war logs. $63\%$ of the
fatalities are labeled as ACF. The second highest fatality number
($16.54\%$) is observed for civilians, closely followed by $15.72\%$
Afghan soldiers and policemen and $1146$ or $4.74\%$ killed allied
soldiers. Palpable are the two peaks for killed insurgents in late
summer 2006 and 2007 in Figure \ref{figdeaths}. They account for
$943$ killed ACF fighters during September $2006$ and for $917$ in
September $2007$. The former peak corresponds to ``Operation Medusa,''
an operation that had the aim to establish government control over
areas of the Kandahar province. The latter marks operations near
Kandahar in an effort to remove insurgents who had returned to this area.
Mid to late $2009$ is the bloodiest period for civilians, coalition
soldiers and ACF in the data. Between May $2009$ and December $2009$ we
observe $1056$ ($26.4\%$) out of $3994$ civilian fatalities (see
Table \ref{tabsumm}). In August $2009$, during the period of the
Afghan presidential election (August 20), we observe $206$ civilian
victims and $190$ killed ACF.
For both groups, this is the highest death toll within one month.
Roughly the same situation is observed for allied soldiers. Here the
monthly maximum of $90$ deaths happens in July $2009$ and from
May $2009$ to December $2009$ the data account for $346$ ($30.2\%$)
killed allied soldiers.\looseness=1

In addition to the fatality numbers, the reports contain $28$ numerical
and factor variables that serve as split candidate variables for the
segmentation. We restrict ourselves to describing only those splitting
variables that have a special relevance for our analysis.

The factor \texttt{attackOn}, with its levels \texttt{FRIEND}, \texttt{NEUTRAL},
\texttt{ENEMY}, \texttt{UNKNOWN} encodes the US military's point of view on whom an
``attack'' (action) is directed during the incident. {O'L}oughlin
et al. [(\citeyear{OLoughlin2010}), page 474, ff] state that this
variable seems to have been mislabeled and should have been named
``attackBy.'' However, after inspection of a random sample of about
$100$ report summaries of the war logs, we believe that \texttt{attackOn}
does not contain information about who carried out a certain action but
rather contains information about on whom the action described in the
report is directed. For instance, leaflets of Anti Coalition Forces
(ACF) calling for attacks against the US forces are categorized as
\texttt{attackOn=NEUTRAL}, fire fights between ACF and allied soldiers
as \texttt{attackOn=ENEMY} and friendly fire is labeled as
\texttt{attackOn=FRIEND}.

The categorical variable \texttt{dcolor} controls the display color of
the message in the messaging system and map views. Messages relating
to enemy activity have the color \texttt{red}, those relating to
friendly activity have been colored \texttt{blue}, and \texttt{green}
stands for neutral. This variable can be seen as the one encoding by
whom an action has been carried out (i.e., ``attackBy'').

Another important variable for our analysis is \texttt{region}, roughly
describing where an event took place. It has levels \texttt{RC NORTH},
\texttt{RC}
\texttt{EAST}, \texttt{RC WEST}, \texttt{RC SOUTH},
\texttt{RC CAPITAL}, \texttt{UNKNOWN} and \texttt{NONE SELECTED}
(RC stands for ``Regional Command'').

Last, there is \texttt{complexAttack}, a categorical variable with
levels \texttt{TRUE}, \texttt{FALSE} and \texttt{NA} (not available)
that encodes the complexity of an attack. The US military states an
attack as complex if it has been well organized and executed, if
soldiers have made use of heavy artillery and the troops have been able
to withdraw from the battlefield in an organized fashion [see \citet{lwj2009}].

\subsection*{The report summaries}
\label{secsummary}
The variables described above, which may serve as split candidate
variables for segmenting the data set, only allow for a rather limited
view into the events associated with each report and therefore the
circumstances under which fatalities have happened. We can, however,
find additional information about the context of the various incidents
in the provided report summaries, which contain a short verbal
description of what transpired during the incident. To give an example,
for 19-Jul-2005 we can find the following report:

\begin{quote}
On 19 July, at about 0730 hrs, a BBIED went off on an alleged suicide
bomber targeting Enjeel district Chief of Police. As a result, the
attacker was instantly killed, but no injures to anyone else was
reported. Police investigation is ongoing.
\end{quote}

\noindent The report summaries tell us the hows and whys of the mission in a very
detailed way, something the other provided variables cannot. Thus, the
report summaries and their content are at the core of evaluating the
ongoings of this war as portrayed in the war logs as well as gaining
insight into mortality in different situations. Disregarding these
summaries in evaluating the war logs would be equivalent to discarding
the most important information for describing under which circumstances
deaths happen.

However, making use of this information is challenging. First, the
summaries are plain natural language text which we need to process.
Second, the bulk of reports makes processing of the summaries by humans
rather difficult. A person would have to read or process more than
$76\mbox{,}900$ texts. If each summary takes a minute to read and file or
process in any way, it would amount to approximately $1282$ hours of
work (or $160$ work days if a work day consists of 8 hours).

There are three possible strategies to deal with such data: either the
reports are processed by crowdsourcing them to a high number of
people. Or, if there is an a priori defined category system,
one may classify the reports into these categories with a supervised
approach. Both strategies were not feasible. Hence, we used a technique
that at the same time generates a category system and provides
meta-information, which
can then be used for aggregating reports with similar content.

\section{Method}
\label{secmethod}
\subsection{Using topic models to build splitting variables from
report summaries}
\label{sectopicmodels}
There exist several approaches for extracting and handling textual
information from documents. One strategy is to cluster the
documents by matching them against predefined queries of terms,
with the drawback that this might be inaccurate due to
polysemy (multiple meanings) and synonymy of single terms. Latent
Semantic Indexing (LSI) [\citet{Deerwester1990}] overcomes this by
performing a singular value decomposition and thus mapping terms and
documents into a latent semantic space. LSI provides more robust
indicators of meaning than simple clustering but lacks in terms of a
solid probabilistic foundation. This is solved by \citet{Hoffman1999}
and his seminal work on probabilistic LSI (pLSI). In pLSI, each word in
a document is modeled as a sample from a mixture model specified via
multinomial random variables. One drawback of pLSI, however, is that it
provides no probabilistic structure at the level of documents. \citet
{Blei2003} fill this gap by the specification of Latent Dirichlet
Allocation (LDA).

LDA is a powerful document generative hierarchical model for clustering
words into topics and documents into mixtures of topics. In LDA the
topics are assumed to be uncorrelated
[but see \citet{blei2007}, for a version with correlated
topics].\vadjust{\goodbreak}
Assuming that the similarity of the circumstances between reports is
reflected in the words contained in the respective summaries, we can
use LDA to assign reports based on their summaries to a number of
topics lower than the number of documents. Hence, in this fashion we
use the allocation of each report to (one or more) latent topic(s) as a
task of complexity reduction or as a preprocessing step.

According to \citet{Blei2009}, topics are automatically discovered
from the original texts and no a priori information about the
existence of a certain theme is required. This means LDA generates the
category system by itself. Only the number of topics for the whole set
of documents has to be specified. The resulting topics are shared
across the whole set of documents. Please note that in general the
topic distribution of each report does only include nonzero probabilities.

Regarding the appropriateness of topic models for such a task, \citet
{Chang2009} presented results of a comparison of topic models with
human classification. They concluded that ``humans are able to
appreciate the semantic coherence of topics and can associate the same
documents with a topic that topic model does'' [\citet{Chang2009},
page 8]. Along similar lines, Griffiths and Steyvers
[(\citeyear{Griffiths2004}), page 5228] note
that ``the extracted topics capture meaningful structure in the data,
consistent with the class designations provided by the authors.''

\subsubsection{The report generative LDA model}
Following \citet{Blei2009} and \citet{blei2011}, LDA specifies the
report generating process as a probabilistic model, in which each
report is a mixture of a set of topics and each word in a report is
chosen from the selected topic specific word distribution.

More formally, let $q$ denote the size of a vocabulary (unique words
within the considered corpus of reports) and let $s$ be the number of
topics $\bolds{\beta}_t, t=1,\ldots,s$. Each topic $\bolds
{\beta}_t$ is a $q$-dimensional symmetric Dirichlet distribution over
the vocabulary with scalar parameter $\eta$. The only observed
variables are words $\mathbf{w}_{1:n}$, where $n$ denotes the
number of reports and $w_{d,m}\in\{1,\ldots,q\}$ denotes the $m$th
word of document $d$. The reports $d$, $d=1,\ldots,n$, are sequences of
those words of varying lengths $q_d$. Each report $d$ is assigned to
topics with the assignments denoted by $z_d$ and the topic assignment
of each of its words $w_{d,m}$ is denoted by $z_{d,m}$. Each report is
seen as a mixture of topics and, hence, it has a vector of topic
proportions denoted by $\bolds{\pi}_d$, with $\pi_{d,t}$
denoting the proportion of topic $t$ in report $d$. The distribution of
$\bolds{\pi}_d$ is an $s$-dimensional symmetric Dirichlet
distribution with scalar parameter $\kappa$.
Hence, the generative model for LDA is
%
\begin{eqnarray}
\label{eqtopic1}\qquad
&&
P(\mathbf{w}_{1:n},\bolds{
\beta}_{1:s},\bolds{\pi }_{1:n},\mathbf{z}_{1:n}
\vert\eta,\kappa)
\nonumber\\[-8pt]\\[-8pt]
&&\qquad=\prod_{t=1}^{s}P(\bolds{
\beta}_t\vert\eta)\prod_{d=1}^{n}
\Biggl[P(\bolds{\pi}_d\vert\kappa) \Biggl(\prod
_{m=1}^{q_d} P(z_{d,m}|\bolds{
\pi}_d)P(w_{d,m}|\bolds{\beta }_{1:s},z_{d,m})
\Biggr) \Biggr],
\nonumber
\end{eqnarray}
where the conditional distributions of the topic assignments and the
words are assumed to be categorical (multinomial with a single trial),
that is, $z_{d,m}\sim\mathrm{Categorical}_{s}(\bolds{\pi}_d)$
and $w_{d,m}\sim\mathrm{Categorical}_{q}(\bolds{\beta
}_{z_{d,m}})$. 
For estimation of the model we employed the variational EM-Algorithm,
which has the effect that $\eta$ can remain unspecified [see, e.g.,
\citet{Gruen2011}]. Since we use LDA to generate topics and assign
each document to one of them, we need the posterior distribution of the
latent topics, the topic assignment and the topic proportions given the
documents,
%
\begin{equation}
P(\bolds{\beta}_{1:s},\bolds{\pi}_{1:n},\mathbf
{z}_{1:n}|\mathbf{w}_{1:n},\eta,\kappa)=\frac{P(\mathbf
{w}_{1:n},\bolds{\beta}_{1:s},\bolds{\pi
}_{1:n},\mathbf{z}_{1:n})}{P(\mathbf{w}_{1:n})}
\end{equation}
and the conditional expectations $\bolds{\hat{\beta
}}_{t,u}=E(\bolds{\beta}_{t,u}|\mathbf{w}_{1:n})$,
$\bolds{\hat{\pi}}_{d,t}=E(\bolds{\pi}_{d,t}|\mathbf
{w}_{1:n})$ as well as $\hat{z}_{d,t}=E(Z_{d}=t|\mathbf
{w}_{1:n})$ with $u=1,\ldots,q$.

We follow suggestions in the pertinent literature [see
\citet{Blei2003,Titov2008,Steyvers2004}] and omit stop words from
the reports. Additionally, we use a stemmer to canonicalize different
inflected forms to their base form (e.g., friends to friend). We then
specify an a priori number of $100$ latent topics to be estimated from
the stop word free corpus of stemmed words. In addition, we set the
parameter $\kappa$ of the symmetric Dirichlet distribution of the topic
proportions to a very small value ($0.001$) in order to ensure that the
estimated topic distribution for each document will assign a
probability of nearly one to a single topic and very small
probabilities to all other topics. This makes it possible to switch
from soft to hard assignments without substantial loss of information.
The resulting dummy variables that encode whether a document belongs to
a topic or not then serve as split candidate variables for subsequent
analysis of the fatality numbers.

\subsection{Recursive partitioning of negative binomial distributions}
\label{secmixture}
Our target variable is the number of fatalities per report $Y_d$
$(d=1,\ldots,n)$, with realizations~$y_d$. We use model trees with a
prespecified node model to segment the data. This allows to incorporate
information from $p$ split candidate variables $\mathbf
{x}_d=(x_{1d},\ldots,x_{pd})^T$ for segmentation. Note that we model
the fatalities locally in each segment and identify the segments based
on statistical inference for the node models. This idea has objectives
similar to model-based clustering. We choose recursive partitioning
rather than mixture models because trees (i) expect all variables to
interact with each other, (ii) automatically detect interactions, (iii)
yield parsimonious interaction patterns, (iv) conduct variable
selection due to the greedy forward search and (v) do not need the
number of segments to be specified a priori.\looseness=1 

More formally, the conditional distribution of $Y$, $D(Y|\cdot)$ is
modeled as a partition function $f$ depending on the state of $p$
splitting vectors (variables), $\mathbf{x}=(x_{1},\ldots,x_{p})$, that is,
%
\begin{equation}
D(Y|\mathbf{x})=D\bigl(Y|f(x_1,\ldots,x_p)\bigr),
\end{equation}
where the function $f$ partitions the overall splitting variable space
$\mathcal{X}$ into a set of $r$ disjoint segments $R_1,\ldots,R_r$ such
that $\mathcal{X}=\bigcup_{k=1}^{r}R_k$ [\citet{hothorn06}]. In each
segment $R_k$, a local model for the conditional distribution is
fitted.

Our model for the conditional distribution $D(Y|\mathbf{x})$ within
each segment $R_k, k=1,\ldots,r$, is a negative binomial distribution
with mean $\mu_k$ and shape parameter $\theta_k$, that is, having the
probability mass function
%
\begin{equation}
\label{eqnegbin} P(Y=y|k;\mu_k,\theta_k)=
\frac{\Gamma(y+\theta_k)}{\Gamma(\theta
_k)y!} \biggl(\frac{\mu_k}{\mu_k+\theta_k} \biggr)^y \biggl(
\frac
{\theta_k}{\mu_k+\theta_k} \biggr)^{\theta_k}
\end{equation}
with $y \in\{0,1,2,\ldots\}$, and $\Gamma(\cdot)$ denoting the gamma
function. Mean and variance of $Y$ for each segment $R_k$ are given by
[\citet{lawless87}]
%
\begin{equation}\label{eqmeanNB}
\mathrm{E}(Y)=\mu_k,\qquad\operatorname{Var}(Y)=\mu_k+
\mu_k^2\theta_k^{-1}
\end{equation}
and the segment size by $n_k$. Please note that the above formulation
pays dues to interpreting the negative binomial as a gamma mixture of
Poisson distributions [\citet{aitkin09}] and thus essentially being a
Poisson model that can account for extra variation. It can be seen as a
two-stage model for the discrete response $Y$ in each segment $R_k$
[cf. \citet{venables02}],
%
\begin{equation}
Y|V \sim\operatorname{Poisson}(\mu_k V),\qquad\theta_k V\sim
\operatorname{Gamma}(\theta_k).
\end{equation}
Here $V$ is an unobserved random variable having a gamma distribution
with mean~$1$ and variance $1/\theta_k$. However, the marginal
mean--variance identities for $Y$ in (\ref{eqmeanNB}) hold whenever
$V$ is a positive-valued random variable with mean $1$ and variance
$\theta_k^{-1}$ and $V$ need not necessarily be gamma
distributed [\citet{lawless87}]. 
Using the negative binomial distribution has the advantage over a
Poisson model to account for extra variation and over Quasi-Poisson to
integrate nicely into a maximum likelihood framework [see
\citet{venables02}]. In principle, the other count data models might
also be used as the node model. In fact, a Quasi-Poisson model tree
approach for modeling overdispersed count data has been proposed by
\citet{choi05}. Their rationale is similar to ours, but we use
negative binomial distributions to account for overdispersion and a
different tree algorithm that is unbiased in variable selection. The
last point is very important for the correct interpretation of the tree
structure [\citet{Loh1997,Loh2002,kim2001}] and depends on the
splitting procedure [\citet{Loh2009}].

\subsubsection{Estimation}
For simultaneous estimation of the segmentation and the node model
parameters, we employ the model-based recursive partitioning framework
of \citet{zeileis08}. Hereby we consider an intercept-only model
(i.e., there are no explanatory variables in the node model) estimated
from a negative binomial likelihood which is then recursively
partitioned based on the state of the split variables. For GLM-type
models such as the negative binomial model, the algorithm is described
in detail in \citet{rusch11}. This algorithm ensures that split
variable selection is practically unbiased.

As tuning parameters for the tree algorithm we have the global
significance level $\alpha$ of the generalized M-fluctuation tests
[\citet{zeileis07}] used for split variable selection and the minimum
number of observations per node. Setting the former to low values can
be regarded as pre-pruning to avoid overfit. As suggested for this
procedure [\citet{zeileis08}], we let qualitative considerations guide
our choice of tuning parameters. For this data set, significance levels
of around $0.01$ or higher might lead to spuriously significant results
due to sample size, hence, we chose a low significance level of
$1\times10^{-4}$. Additionally, we wanted to have at least $0.4\%$ of
the overall observations in a segment. Both choices were made to reduce
fragmentation of the tree and to get a number of segments somewhere
between 10 and 20.

Eventually we get a classification of all observations into a set of
segments $\mathcal{R}=\{R_1,\ldots,R_r\}$. The negative binomial
distributions in these segments are characterized by the parameter
estimates $\hat{\mu}_k$ and $\hat{\theta}_k, k=1,\ldots,r$, and the
estimated overall tree model by $\hat{\bolds\vartheta}=((\hat
{\mu}_1,\hat{\theta}_1)^T,\ldots,(\hat{\mu}_r,\hat{\theta}_r)^T)$.

\subsubsection{Interpretation of the models}
Basically, interpretation happens on two levels: first, the level of
the individual segments for which we get the estimated mean number of
fatalities as well as the associated standard deviation. These fatality
rates identify which segments come along with a higher or lower average
death toll. Second, the level of the splitting variables that define
the segments. Here conclusions can be drawn about the specific
circumstances that give rise to the segments and hence to the different
fatality rates. In the case of topics as splitting variables, we only
look at which topics are selected for splitting and interpret them
\textit{ex post} based on their most frequent words. Hence, topics are
used only for splitting without any further interpretation of or prior
hypothesis about the underlying topic model. For readability we assign
a unique name to each topic, but it should be kept in mind that those
names are somewhat arbitrary. Since they are derived solely from the
ten most frequent words as well as from looking at a random sample of
assigned report summaries, they are necessarily neither exhaustive in
their denotative and connotative meaning nor can they capture the
circumstantial complexity of all assigned reports.

\section{Results and discussion}
\label{secresults}
In our analysis the response was the overall fatality number (sum of
fatalities of civilians, the ACF, of coalition troops and of Afghan
police and soldiers). Detailed analyses for all groups separately can
be found in \citet{resrep}.

\begin{sidewaysfigure}

\includegraphics{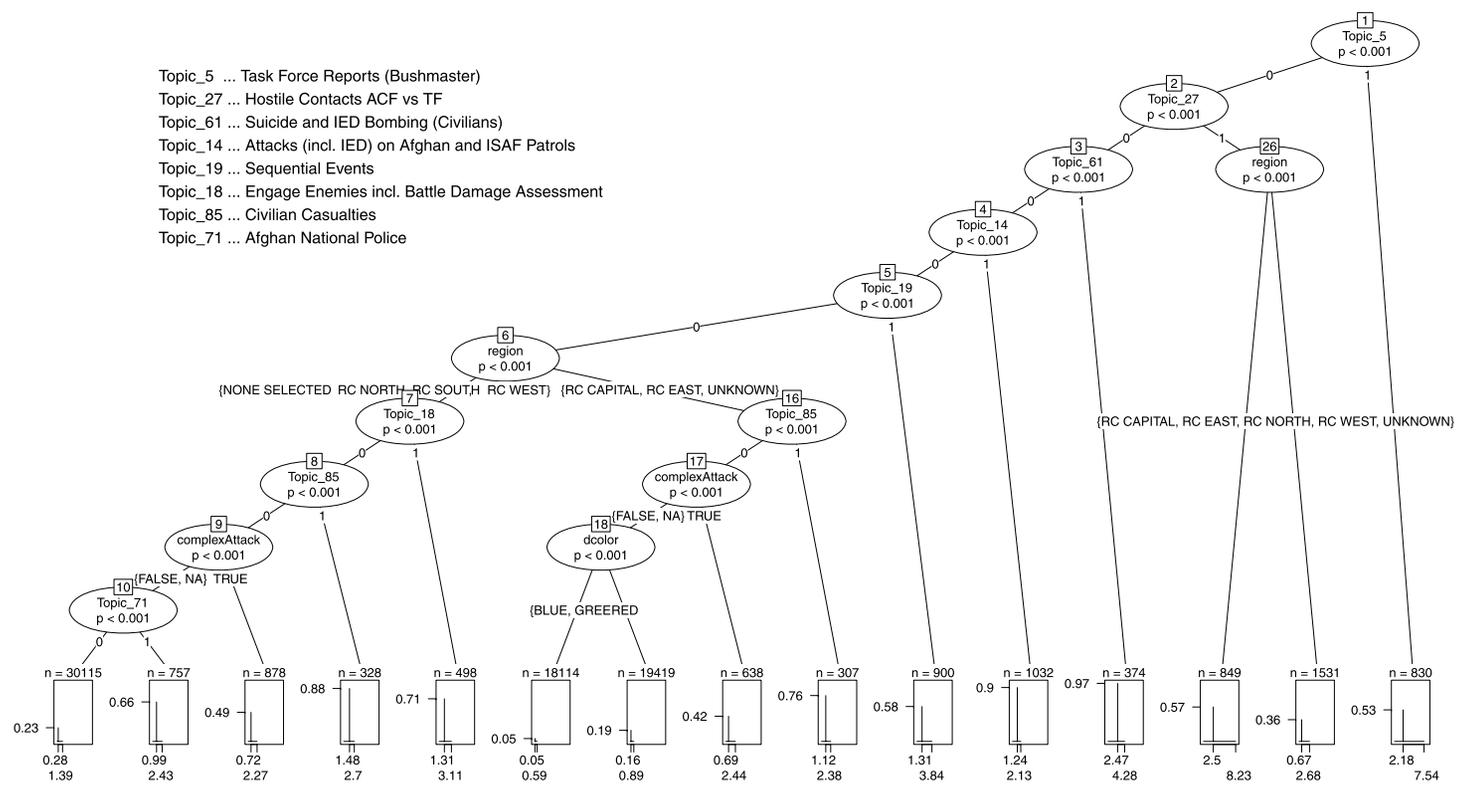}

\caption{The negative binomial model tree for the combined fatalities.
In the segments the vertical line marks the mean, the horizontal line
the length between zero and one standard deviation and the height of
the vertical line is the deviance divided by the degrees of freedom (we
included a larger version of this plot in
\protect\hyperref[suppA]{Supplement A}
[Rusch et~al. (\citeyear{supplementa})]).} \label{figall}
\end{sidewaysfigure}

Along the lines of the methodological procedure described above and to
understand the fatality numbers associated with different
circumstances, we first need the split information, that is, which
topics or further variables have been selected as splitting variables
as well as where the split occurred. Second, we need the estimated
parameters of the segment-specific model, that is, mean and shape.
Accordingly, the split information is presented in Figure \ref
{figall} and the estimated node model parameters in Table \ref{tblresAll}.

\begin{table}
\caption{Segment-wise statistics for all fatalities combined. The
first column refers to the segment. For each segment we listed the
logarithm of the estimated mean $\log(\hat{\mu}_k)$, its
standard\vspace*{1pt}
error $\operatorname{se}(\log(\hat{\mu}_k))$, the estimated shape parameter
$\hat{\theta}_k$ and its standard error $\operatorname{se}(\hat{\theta
}_k)$, the degrees of freedom ($df_k = n_k-1$), the residual deviance
(dev), the highest number of fatalities reported (max) and the
percentage of reports with zero fatalities ($\%$zero)}
\label{tblresAll}
\begin{tabular*}{\tablewidth}{@{\extracolsep{\fill}}ld{2.3}cccrrd{3.0}r@{}}
\hline
\textbf{Segment}&\multicolumn{1}{c}{$\bolds{\log(\hat{\mu}_k)}$}
&\multicolumn{1}{c}{$\bolds{\operatorname{se}(\log(\hat{\mu}_k))}$}
& \multicolumn{1}{c}{$\bolds{\hat{\theta}_k}$}
& \multicolumn{1}{c}{$\bolds{\operatorname{se}(\hat{\theta}_k)}$}
& \multicolumn{1}{c}{$\bolds{df_k}$}
& \multicolumn{1}{c}{\textbf{dev}} & \multicolumn{1}{c}{\textbf{max}}
& \multicolumn{1}{c@{}}{\textbf{\%zero}}\\
\hline
$R_1$ &0.779 &0.120&0.089&0.007&829&436.36&101&$75.4$\\
$R_2$ &-0.399&0.102&0.069&0.006&1530&554.37&68&$84.8$\\
$R_3$ &0.917 &0.113&0.096&0.008&848&486.90&186&$72.4$\\
$R_4$ &0.904 &0.090&0.386&0.038&373&361.19&36&$42.8$\\
$R_5$ &0.215 &0.053&0.468&0.037&1031&926.77&31&$53.8$\\
$R_6$ &0.269 &0.098&0.128&0.011&899&523.48&70&$73.1$\\
$R_7$ &0.114 &0.121&0.275&0.039&306&234.08&43&$63.2$\\
$R_8$ &-0.376&0.146&0.090&0.011&637&267.76&25&$82.3$ \\
$R_9$ &-1.804&0.039&0.043&0.002&19\mbox{,}418&3604.80&28&$93.4$\\
$R_{10}$&-2.979&0.086&0.009&0.001&18\mbox{,}113&860.49&67&$98.3$ \\
$R_{11}$&0.269 &0.106&0.205&0.022&497&353.50&56&$66.3$\\
$R_{12}$&0.389 &0.101&0.373&0.046&327&288.75&35&$52.7$\\
$R_{13}$&-0.329&0.106&0.117&0.012&877&425.75&35&$79.3$\\
$R_{14}$&-0.011&0.089&0.199&0.019&756&497.98&21&$70.2$\\
$R_{15}$&-1.282&0.029&0.047&0.001&30\mbox{,}114&6884&80&$91.3$\\
\hline
\end{tabular*}
\end{table}

Regarding splits in the tree based on estimated latent topics, a
presentation of their ten most frequent keywords and how many reports
were assigned to them can be found in Table \ref{tblterms} in
Appendix \ref{app1} along with the absolute word frequency as a
measure of word importance. For instance, the report summary from
Section \ref{secsummary} belongs to Topic 61, ``Suicide and IED
Bombing.'' In Table \ref{tblterms} the ten most frequent words of
this (and other tree topics) are displayed, with ``suicid'' having
occurred 520 times. Additionally, we can see in the first row of
Table \ref{tblterms} (\texttt{numberDOC}) that overall $378$ incidents
were assigned to this topic.

In Figure \ref{figall} we visualize the negative binomial
distribution in each segment by a parsimonious plot of the magnitudes
of the mean and the standard deviation. The vertical line in each panel
marks the location of the mean, the horizontal line shows the distance
between zero and one theoretical standard deviation [cf. \citet
{friendly2001}]. The height of the vertical line is the deviance
divided by the degrees of freedom and indicates fit of the
intercept-only model in the node. A~smaller height means less
dispersion and thus better fit (see also \hyperref[suppB]{Supplement B}
[\citet{supplementb}]).

We labeled the segments $k=1,\ldots,r$ in an increasing order from
right to left as they are displayed in the plot. This is of course
arbitrary and should not imply a natural ordering of the $k$ segments.
Each segment $R_k$ is associated with a local negative binomial
distribution with parameter estimates $\hat{\mu}_k$ and $\hat{\theta
}_k$. For each segment, Table \ref{tblresAll} lists the segment
number, parameter estimates and standard errors, degrees of freedom
$(n_k-1)$, deviance, the maximum number of fatalities and the
percentage of incidents with no fatalities.

In what follows we discuss the results for the most interesting
segments in more detail.

\subsection{Fatalities in the war logs}
\label{secresults1}
For all fatalities combined, we find $r=15$ segments (with a global
significance level for the fluctuation tests of $\alpha=1\times
10^{-4}$ and a minimum number of observations in each segment of
$300$). The resulting tree is depicted in Figure \ref{figall}.

The tree for the overall number of fatalities is dominated by
fatalities of the ACF and of the civilian population. The tree itself
is largely a combination of the trees for ACF and civilian fatalities
alone [see \citet{resrep}]. Our presentation will therefore mainly focus
on ACF fatalities and civilian deaths, since those groups account for
the highest number of deaths. Fatalities of allied forces and the
troops of the host nation play a minor role for the overall number of
deaths due to the comparatively small number of those fatalities
(especially of allied forces) and the high congruency of civilian
deaths and deaths of host nation troops.\footnote{For what follows, it
should be noted that the entries in the database can be prone to data
entry errors, mainly misclassification of fatalities to their
respective group. For instance, the Kunduz air strike incident on
03-Mar-2009 lists $56$ fatalities. All fatalities are stated to be
``ACF fighters'' in the war log. In the media, however, the killed
people were identified as being civilians [see \citet{guardianKunduz}]
who were invited by the Taliban to take fuel from stolen fuel trucks
[see \citet{amnestyKunduz2009}]. An allied air strike against the
fuel trucks killed those 56 civilians. This should be kept in mind,
although generally there is a high congruency between the data in the
WikiLeaks war log and other independent data sets
[\citet{Bohannon2011}].}

The first three segments are dominated by reports listing high numbers
of fatalities of the ACF. These reports belong either to ``Task Force
Reports (Bushmaster)'' or are associated with incidents attributable to
``Hostile Contacts ACF vs TF'' in the South and elsewhere.

The first segment consists of $n_{1}=830$ incidents, with a maximum
number of deaths of $101$. $75.4\%$ of the documents reported no
fatalities. The average fatality number per report for this segment was
$\hat{\mu}_{1}=2.18$ ($2.1$ for ACF alone). The $101$ ACF deaths that
mark the maximum death toll in this segment is the third highest death
number in the whole war diary, as is the mean fatality rate. All in
all, $1808$ deaths are reported in this segment, $1712$ of those are
categorized as ACF. This segment is characterized by reports that
belong to Topic $5$ ``Task Force Reports (Bushmaster).'' Table \ref
{tblterms} displays the most frequent words in the summaries of this
and subsequent topics along with their frequencies. For Topic 5 they
were ``task force,'' ``fire,'' ``close,'' ``track,'' ``insurgencies,''
``bushmaster'' and ``isaf.'' Inspection of report summaries from this
topic suggests that this segment refers to reports by US task forces
(TF) with a focus on actions of task force unit ``Bushmaster.'' TF
``Bushmaster'' is a task force consisting of Afghans and American green
beret soldiers, the latter being a synonym for the United States Army
Special Forces. According to Wikipedia, they have ``six primary
missions: unconventional warfare, foreign internal defense, special
reconnaissance, direct action, hostage rescue, and counter-terrorism.
The first two emphasize language, cultural, and training skills in
working with foreign troops. Other duties include combat search and
rescue (CSAR), security assistance, peacekeeping, humanitarian
assistance, humanitarian de-mining, counter-proliferation,
psychological operations, manhunts, and counter-drug operations''
[\citet{wikiusaf}]. The topic mainly describes events or fights
connected with this and other TF, including detention of individuals,
fights and espionage.

The next two segments are governed by Topic $27$ ``Hostile Contacts ACF
vs TF'' and differ in terms of the region they took place. They
describe incidents where task forces or ground troops had enemy contact
in fire fights taking place (individual combat with small arms, see
Table \ref{tblterms}). Excluded from this topic are reports from
Topic $5$. Incidents assigned to this topic are further split according
to the region where the events took place. The right branch in
Figure \ref{figall} contains events around Kabul (\texttt{RC
CAPITAL}), \texttt{RC EAST}, \texttt{RC WEST}, \texttt{RC NORTH} and
\texttt{UNKNOWN} regions, as collected in segment $R_{2}$ which might be
called ``Hostile Contact ACF vs TF (not in the South).'' These are
associated with a death rate of $\hat{\mu}_{2}=0.671$ ($0.6$ for ACF
alone). Of these $1531$ incidents the maximum number of fatalities is
$68$ and $84.8\%$ reported no fatalities.

Of the reports belonging to Topic $27$ ``Hostile Contact ACF vs TF,''
the $849$ events that happened in the South of Afghanistan (mainly\vadjust{\goodbreak}
provinces Kandahar and Helmand, \texttt{RC SOUTH}) show a much higher
estimated fatality rate of $\hat{\mu}_{3}=2.501$ ($2.4$ for the ACF
alone). This is the highest estimated death rate of any segment. It can
be explained by the South, especially the province of Kandahar, being
Taliban heartland and their stronghold. It is therefore heavily
attacked by coalition troops [see \citet{OLoughlin2010}]. This result
of higher death rates for incidents happening in the South is recurrent
for all groups of fatalities [see \citet{resrep}]. The segment
``Hostile Contact ACF vs TF (South)'' contains, among others, events
that took place during Canadian-led ``Operation Medusa,'' which began
on September 2, 2006 and lasted until September 17 [see \citet
{WikiMedusa2006}]. Reports in this segment ($R_3$) have a maximum
number of fatalities of $186$ on September 9, 2006. This report (its
incident being part of ``Operation Medusa'') notes $181$ killed ACF
fighters, one killed coalition force soldier and four killed Afghan
soldiers $10$ km southwest of Patrol Base Wilson, in Kandahar
province's volatile Zhari district. This is the highest number of
killed ACF fighters (or overall death) within a single war log entry in
the whole data set. Moreover, segment $R_3$ is generally the segment
with the highest ACF fatalities. Still, for $72.4\%$ of the documents
in this segment no fatalities are reported.

The next three segments we discuss consist of incidents that are
characterized by a high death toll of the civilian population mainly
resulting from actions of the ACF.

First, there is Topic $61$ ``Suicide and IED Bombing'' with
corresponding segment $R_{4}$. It describes incidents that were related
to suicide bombing attacks or other attacks with improvised explosive
devices (IED) such as cars (cf. Table \ref{tblterms}). For example,
one report assigned to Topic $61$ and dated with 18-Feb-2008 reports
$30$ killed civilians due to a suicide bomb attack near Kandahar. It
also includes reports where explosives were found or seized. The
segment's $n_4=374$ reports list fatalities in $57.2\%$ of the cases,
which makes it the only segment with a median death number higher than
$0$. The maximum number of killed people is $36$. Accordingly, the
estimated mean death rate for this segment is $\hat{\mu}_{4}=2.471$
($1.12$ for civilians alone, the second highest civilian fatality
rate). It is the second highest overall death rate per incident,
closely matching the results from $R_{3}$. However, in $R_4$ ``Suicide
and IED Bombing'' fatalities are mostly civilians or Afghan police
forces, whereas deaths in $R_{3}$ ``Hostile Contacts ACF vs TF
(South)'' are mostly ACF fighters. In $R_4$ we observe $924$ deaths, of
which $420$ are civilian, followed by $246$ killed afghan soldiers and
$233$ killed ACF fighters.\looseness=1

The next segment is $R_{7}$ ``Civilian Casualties (East, Capital and
unknown regions)'' with an overall average number of fatalities of
$\hat{\mu}_7=1.12$. These are those $n_7=307$ incidents in the East,
the capital or unknown\vadjust{\goodbreak} region associated with Topic $85$ ``Civilian
Casualties.'' In Table \ref{tblterms} we see the clear context of
civilian fatalities of this topic. Out of the ten most frequent terms
of this topic, six are synonyms, respectively, acronyms of civilians.
These are as follows: ``ln'' (local national), ``local(s),''
``civilian,'' ``lns'' (local nationals), ``child,'' ``nationals.'' The
other four terms suggest a clear connection to casualties, namely,
``wound,'' ``injur'' (injury), ``kill,'' ``hospit'' (hospital). The
maximum number of fatalities in this segment is $43$ and there are
$63.2\%$ of reports that list no fatality at all.

Segment $R_{12}$ (governed by events from Topic $85$ ``Civilian
Casualties'' happening in the South, North, West or in a nonspecified
region) has an estimated mean of $\hat{\mu}_{12}=1.476$. The
percentage of reports without killings is $52.7\%$ and the highest
death toll is $35$. The governing topic, Topic $85$, appeared before as
the governing topic of $R_7$. Therefore, $R_{12}$ and $R_{7}$ are
corresponding topic-wise and only differ in terms of their location. It
is interesting to see that $R_{12}$ has a higher fatality number per
incident, most probably due to events in the south. Incidents in Kabul
and the East ($R_{7}$) are associated with lower death numbers and a
higher percentage of reports with zero deaths. However, the report with
the highest fatality number for this topic is part of $R_7$, describing
an attack on the Indian embassy in Kabul where $42$ civilians and one
Taliban were killed.

When looking at civilian fatalities alone, incidents from Topic $85$
``Civilian Casualties'' have the overall highest observed civilian
death toll for actions of the ACF, either against civilians or where
civilians are ``collateral damage'' (on average $1.7$ deaths per
incident). Hence, incidents from this topic as well as incidents in
Topic $61$ ``Suicide and IED Bombing'' have in common that the attacks
were overwhelmingly carried out by the ACF and were directed at places
where there is a high number of the civilian population present, such
as buses, bazars or markets. In contrast, for incidents which refer to
actions of ISAF troops also belonging to Topic $85$ ``Civilian
Casualties,'' we have about $25\%$ of the former rate ($0.41$ deaths
per incident, the fourth highest overall rate for civilians). Thus, ACF
action is associated\vadjust{\goodbreak} with a fourfold increase in expected civilian
fatalities for reports belonging to this topic. It is a clear and
consistent finding that actions of the ACF come along with a higher
civilian death toll than actions of the allied forces. Generally, when
analyzing civilian fatalities alone, most resulting segments with high
civilian fatality rates have in common that they are connected to
attacks by the ACF often with improvised explosive devices [see also
\citet{Bohannon2011}].\looseness=1

The last segment we discuss is governed by Topic $14$ ``Attacks
(incl. IED) on Afghan and ISAF patrols,'' which gives rise to segment
$R_{5}$ with an average number of deaths per incident of
$\hat{\mu}_{5}=1.241$ ($0.32$ for the civilian population and $0.51$
for Afghan troops). In total, we observe $1287$ deaths in the
$n_5=1032$ reports ($53.8\%$ of which had no deaths reported) in this
segment. It is somewhat hard to identify the governing topic with a
unique theme like before, but inspecting a sample of report summaries
indicates that this topic collects reports which describe explosions
of IED or smaller fights or incidents following attacks by the ACF
mainly with Afghan and some ISAF forces that were patrolling,
resulting in battle damage assessment (bda) and medical evacuation. Most
victims in this segment are therefore Afghan soldiers ($529$), but we
also observe $326, 170$ and $262$ killed civilians, ACF and allied
soldiers, respectively.

It should also be noted (and this finding is consistent throughout all
the fatality groups) that segments containing by far the largest number
of reports have on average relatively low death rates per incident and
feature underdispersion. For all fatalities, these are segments
$R_{15}$, $R_{10}$ and $R_9$ with $\hat{\mu}_{15}=0.28$, $\hat{\mu
}_{10}=0.05$ and $\hat{\mu}_{9}=0.16$. They contain $88.35\%$ of all
reports. Hence, most of the everyday happenings in this war come along
with a low death toll. Only in the case of certain events this number
increases. This increase is mainly connected to either fights between
allied forces and the Taliban and other ACF groups (leading to high ACF
fatality numbers) or is characterized by attacks by the ACF who aim at
or tolerate civilian casualties (leading to high civilian or Afghan
troop fatality numbers).\looseness=1

\section{Model validation}
\label{secval1}
To keep in line with the objectives of our model tree approach, we
validate the clustering structure and---locally for each cluster---the
parametric model. Specifically, we (i) assess stability of the tree
structure and reproducibility of the resulting segmentation and (ii)
evaluate the fit of the local models. A detailed exposition of the
validation results is available as \hyperref[suppB]{Supplement~B} [\citet{supplementb}].

\subsection*{Stability of tree structure and segmentation}
We use resampling with replacement to generate data sets of $5/6$ the
size of the original data set. We fit model trees to the resampled data
sets (tuning parameters modified to $\alpha=10^{-3}$ and a minimum
number of observations of $250$-due to the reduction in sample size).
We use two resampling schemes: (i) regular resampling (RRS; i.e.,
drawing data sets of size $5/6 \times n$ by random resampling with
replacement) and (ii) stratified resampling (SRS; i.e., drawing from
each segment $R_k$ a proportion of $5/6 \times n_k$ reports by random
resampling with replacement). For each resampled data set, the fitted
tree is then used to predict the segment and the fatality number for
the reports not part of the set of resampled reports (the out-of-bag
observations). Thus, we get segment assignments for all in-bag and
out-of-bag reports. The procedure is repeated $200$ times per
resampling scheme.

We use a segment-wise version of the Jaccard index
[\citet{jaccard1901}] as the measure of segment stability and report
concordance. See \hyperref[suppB]{Supplement B} or \citet{Hennig2007} for details.
Possible alternative measures include ``prediction strength''
[\citet{tibshirani2005}]. Let $T$ denote the original tree and $T^{(b)}$
the tree fitted on bootstrap sample $b$ with $T$ having the segments
$R_k, k=1,\ldots,r$, and $T^{(b)}$ the segments $R_l^{(b)},
l=1,\ldots,r^{(b)}$. We denote the segment-wise Jaccard index for each
resample $b$ by $\mathrm{Jac}^{(b)}_{kl}$ with $k=1,\ldots,r$ and
$l=1,\ldots,r^{(b)}$. For each resample $b$ and given segment $R_k$ we
calculate the segment-wise indices $\mathrm{Jac}^{(b)}_{kl}$ and assign the
segment $R^{(b)}_l$, $l\dvtx\arg\max_{l}\mathrm{Jac}^{(b)}_{kl}$ to be the
corresponding segment of $R_{k}$, that is, with concordance
$\mathrm{Jac}^{*(b)}_{k}=\max_{l} \mathrm{Jac}^{(b)}_{kl}$ (see \hyperref[suppB]{Supplement B} for details).

\begin{figure}

\includegraphics{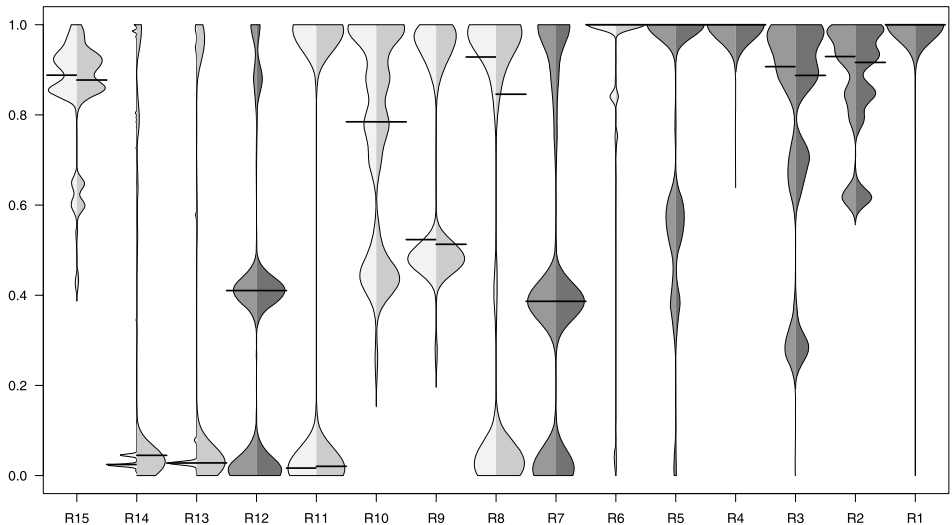}

\caption{Bean plots of the segment-wise Jaccard indices,
$\mathrm{Jac}^{*(b)}_{k}$, between original segments and the matched segments
over the bootstrap samples $b$ for all $R_{k}$, $k=1,\ldots,15$. Darker
beans mark segments we described in detail in Section \protect\ref
{secresults1}. The left part of each bean is a kernel density estimate
for RRS (slightly lighter shaded) and the right-hand side for SRS (slightly
darker shaded). The solid black lines are the medians.}
\label{figjaccardbean}
\end{figure}

When investigating the corresponding tree segments, we find that pooled
over the RRS and SRS scheme (the results do not differ much for each
scheme, see Figure \ref{figjaccardbean} and the supplementary material
[Rusch et~al. (\citeyear{supplementa,supplementb})]) there are $27.6\%$ coinciding segments
($\mathrm{Jac}^{*(b)}_k=1$) and $56.3\%$ strongly corresponding segments
($\mathrm{Jac}^{*(b)}_k \geq0.8$) over the 400 resamples.

This is more pronounced for the segments discussed in detail in
Section \ref{secresults1}. Here we have $42.3\%$ coinciding segments
and $62.4\%$ of the segments show strong correspondence. Note that the
first five described segments, $R_1$ through $R_5$, show even higher
frequencies ($56.2\%$ coinciding and $79\%$ strongly corresponding).
Thus, the results can be considered to be stable with the exception of
the segments associated with Topic 85 ($R_7$ and $R_{12}$), which have
a percentage of $9.2\%$ coinciding segments and $20.9\%$ strongly
corresponding segments.

Regarding stability of the individual segments, the distribution of
the concordance measure $\mathrm{Jac}^{*(b)}_{k}$ for each segment $R_k$ over
the bootstrap samples is summarized with
bean plots [\citet{kampstra2008}] in Figure \ref{figjaccardbean}. The
solid black horizontal lines
denote the medians.
High stability (median${}\geq{}$0.79) is given for 9 out of
15~segments: $R_1$ through $R_{5}$ (from Section \ref{secresults1})
as well as $R_6$, $R_8$, $R_{10}$ and $R_{15}$. For those, $50\%$ of
the corresponding segments\vadjust{\goodbreak} show a concordance of at least $0.79$. The
mass of the Jaccard values is usually concentrated near the median, the
exceptions being $R_{10}$ and $R_8$ and to a minor degree $R_{2}$ and $R_3$.
For certain segments variability is quite high, particularly for
$R_{10}$ and $R_8$. For the segments discussed in detail in
Section \ref{secresults1} the stability is highest, with a median of
$0.79$ or higher in 5 of 7 segments. Low stability is found for
segments $R_{14}$, $R_{13}$ and $R_{11}$. Also, $R_{12}$ and $R_{7}$ are
not particularly stable.

\subsection*{Segment-wise variability of fatality rates}
\begin{figure}

\includegraphics{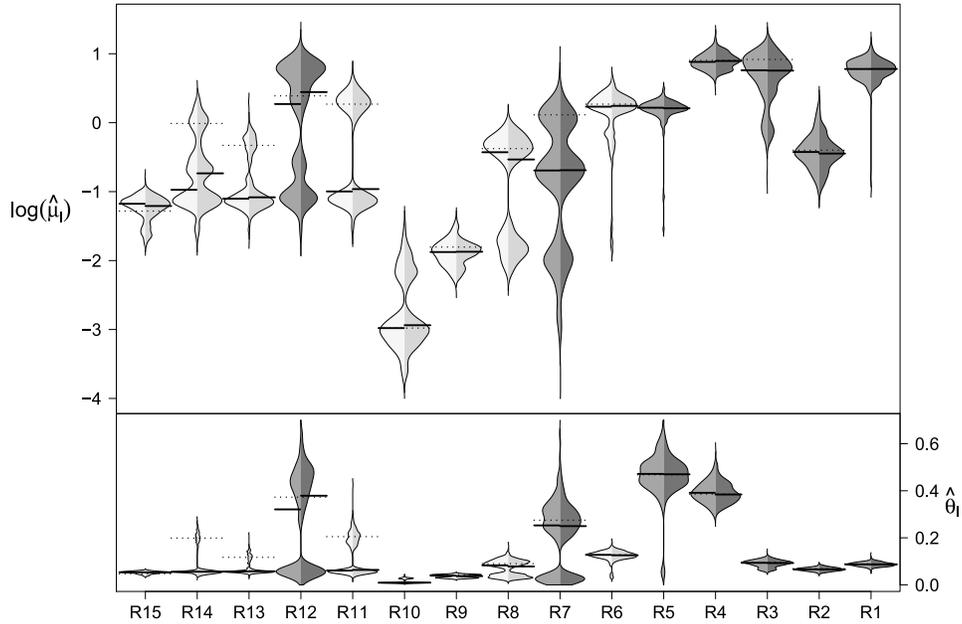}

\caption{Bean plots of the segment-wise estimated death toll parameter
$\log(\hat{\mu}_l)$ and shape parameter $\hat{\theta}_l$ for the
matched segments. The dotted horizontal lines indicate the values of
the original tree (compare Table \protect\ref{tblresAll}). Again, darker
beans mark segments described in detail in the paper. The left part of
each bean is for RRS (slightly lighter shaded) and the right-hand side for
the SRS (slightly darker shaded).}
\label{figmeanbean}
\end{figure}

To evaluate stability of the local models, we investigate the
variability of the estimates of the model parameters for each segment.
We match a given segment $R_k$ from $T$ with a segment $R^{(b)}_l$ from
$T^{(b)}$ based on the highest Jaccard index for each bootstrap sample
as before. For each segment $k$, Figure \ref{figmeanbean} displays
bean plots of the distributions of the parameter estimates $\log(\hat
{\mu}_l)$ and $\hat{\theta}_l$ for the matched segments over all
samples. The dotted horizontal lines indicate the parameter values
estimated for the original tree. The results are in line with those
presented before. We have ten stable segments of the original tree
$R_1$ through $R_6$, $R_{8}$ through $R_{10}$ and $R_{15}$. Over the
bootstrap samples, their median estimated parameter values in the
segments turn out to lie close to the original $\log(\hat{\mu}_k)$.
For most $R_k$ the variability of $\log(\hat{\mu}_l)$ over the
bootstrap samples is rather small. Among them are five segments that we
described in detail in Section \ref{secresults1}, associated with the
topics ``Task Force Reports (Bushmaster),'' ``Hostile Contacts ACF vs
TF,'' ``Suicide and IED Bombing'' and ``Attacks (incl. IED) on Afghan
and ISAF patrols.'' They are reproducible both in terms of the assigned
reports and the parameters for the local models. We also have three
unstable segments ($R_{11}$, $R_{14}$ and $R_{13}$) and two low to
moderately stable segments ($R_{12}$ and~$R_7$). They are practically
the same segments that turned out to be unstable in the previous
section. These segments appear further down the tree hierarchy (see
Figure \ref{figall}) and arise from the branches after the split of
node 6 based on \texttt{region}. Among them are the segments
associated with Topic 85 ``Civilian Casualties'' ($R_7$~and $R_{12}$).
To have a more reliable description for these two segments, an analysis
considering only the number of civilian fatalities might be better and
can be found in \citet{resrep}.

%

\subsection*{Appropriateness of the node model}
To judge the fit of the local models in the nodes, we report the
deviance and the degrees of freedom in Table~\ref{tblresAll}. As can
be seen in detail in Section 2.1 of \hyperref[suppB]{Supplement B}, the deviance values,
their ratio to the degrees of freedom, the mean absolute prediction
error and the residuals all point to a good fit of the segments-wise
models (although for some segments we find substantially less
variability as the model would predict). We further compared the fit of
the negative binomial model to alternative count data models per
segment. Each segment shows substantial overdispersion as compared to a
Poisson model (see Section 2.2 in \hyperref[suppB]{Supplement B}). Inflation of zeros for
a negative binomial model could not be found in any segment. In each
segment the negative binomial model was the count data distribution
with lowest AIC/BIC and highest likelihood (see Section 2.3 in
\hyperref[suppB]{Supplement B}). We also checked for severe violations of the temporal
independence assumption for residuals in each segment. We generally
find no to small autocorrelation in the order the reports have been
filed (see Section 2.4 in \hyperref[suppB]{Supplement B}), hence, the independence
assumption appears to be an acceptable approximation.

\section{Conclusions}
\label{secConclusion}
Undoubtedly, innovations like the internet have changed the supply of
potential data of interest. For science as well as journalism, it is
unavoidable to gather, manage and process this bulk of information.
Central to this is reading, interpreting and understanding text
documents with the aid of automated procedures. The foreseeable
increase of available written information, for example, in the world
wide web, will even increase the need for such methods. At least
partly, this has nourished data journalism and computational social
science where complex data sets become the center of journalistic and
scientific work. This paper illustrates how modern statistical
procedures can provide aid in extracting relevant information from
bulks of written text documents or from a database and how they may
help in processing and structuring the information to facilitate
interpretation of the data, as has been the primary goal of statistical
modeling ever since.\looseness=1

Text mining tools and topic models were used to analyze written text
from the WikiLeaks war diary automatically by assigning overarching
themes to the single documents. This allowed to get a view on the data
which is hard to obtain by manual processing and that may even
discover connections between documents which may not be at all
obvious. The assignment of topics to the single documents offered the
opportunity to use those topics as splitting variables in further data
analysis. One has to bear in mind, however, that the assignment of
documents to topics is by far not absolute and that it can be
difficult to interpret the meaning of latent topics, especially if
they are to be named (as is often the case with unsupervised
techniques). At any rate, we saw that split candidate variables
generated by preprocessing with LDA proved to be very important in the
subsequent analysis, whereas the variables that were already available
played a minor role. Hence, discarding the information stored in the
report summaries would have led to completely different segmentation,
description and interpretation.

Model-based trees were then used to find segments in the data as well
as for providing an intuitive association of circumstances and
fatalities. A representative local data model (here the negative
binomial distribution) was used to relate the observations to the
question at hand. Instead of simply calculating the arithmetic mean of
the dependent variable, the underlying model takes a whole likelihood
for overdispersed count data---suitable for the description of rare
events---into account when estimating the segmentation, the mean
fatality rate and the variance in each segment. Pre-pruning with an
inferential splitting procedure led to a segmentation that proved to be
rather stable in resampling experiments, especially with respect to the
segments that we primarily focused on. The segment assignment of
reports and the estimated parameter values were reproducible when
applied to random subsets of all reports. The local models in the
segments fit the data at hand well. The model tree based segmentation
approach that we chose therefore offered reliable, additional insight
into what the fatality rates for specific incidents look like,
something that has not been done so far for this war.

This clearly illustrates the high potential that text mining
procedures, on the one hand, and model-based recursive partitioning, on
the other, have for a wide range of possible applications in social
sciences [see, e.g., \citet{Kopf2010}] as well as data journalism,
especially if the data stem from a database or consist of both
numerical variables and written text which has to be analyzed, for
example, with data from online forums, social media or social networks.

Despite the insights our approach can provide, we see room for
improving it in future research. First, we did not exploit all of the
spatial and temporal information that is contained in the data set.
While revising this paper, we became aware of the work by \citet
{zammit2012} who made use of the temporal and spatial aspects. It
might be interesting to combine their and our strategy by using their
model as a node model and partition it based on the generated topic
assignments. Second, instead of a two-step procedure, we started
working on a generic model that includes both the preprocessing step as
well as the step of fitting the count data model
simultaneously.\looseness=1\footnote{We thank an anonymous reviewer for this suggestion.}

\begin{appendix}\label{app}
\section{Frequent terms of the topics}
\label{app1}
In Table \ref{tblterms} a list of the ten most frequent terms for
each topic as well as their occurrence for different fatality groups
and the number of documents assigned to them can be found.

\begin{sidewaystable}
\tabcolsep=0pt
\textwidth=\textheight
\tablewidth=\textwidth
\caption{The ten most frequent terms of the estimated latent topics
and the number of documents assigned. A~$\times$ denotes that this
topic serves as a split variable for the mentioned subgroup as
well. Numbers in brackets indicate the term frequencies in the assigned
reports}
\label{tblterms}
\begin{tabular*}{\tablewidth}{@{\extracolsep{\fill}}lrrrrrrrr@{}}
\hline
& \multicolumn{1}{c}{\textbf{Topic 5}}
& \multicolumn{1}{c}{\textbf{Topic 14}} & \multicolumn{1}{c}{\textbf{Topic 18}}
& \multicolumn{1}{c}{\textbf{Topic 19}} & \multicolumn{1}{c}{\textbf{Topic 27}}
& \multicolumn{1}{c}{\textbf{Topic 61}} & \multicolumn{1}{c}{\textbf{Topic 71}}
& \multicolumn{1}{c@{}}{\textbf{Topic 85}} \\
\hline
numberDOC & \multicolumn{1}{c}{830} & \multicolumn{1}{c}{1035}
& \multicolumn{1}{c}{508} & \multicolumn{1}{c}{900} & \multicolumn{1}{c}{2382}
& \multicolumn{1}{c}{378} & \multicolumn{1}{c}{1288} & \multicolumn{1}{c@{}}{638} \\
CIVILIAN & & \multicolumn{1}{c}{$\times$} & & & & \multicolumn{1}{c}{$\times$} & \multicolumn{1}{c}{$\times$} & \multicolumn{1}{c@{}}{$\times$} \\
ACF & \multicolumn{1}{c}{$\times$} & & \multicolumn{1}{c}{$\times$} & \multicolumn{1}{c}{$\times$} & \multicolumn{1}{c}{$\times$} & & & \\
ISAF & & \multicolumn{1}{c}{$\times$} & & & & \multicolumn{1}{c}{$\times$} & & \\
HOST & & \multicolumn{1}{c}{$\times$} & & & & \multicolumn{1}{c}{$\times$} & \multicolumn{1}{c}{$\times$} & \\
[9pt]
& tf (1994) & wia (2004) & engag (1761) & updat (4157)& fire (3345)
&suicid (520) & anp (4132) & ln (1595)\\
& bushmast (1570) & ie (1981) & fire (1092) & att (1742) & tf (2979)
& bomber (470) & event (643) & wound (799) \\
& fire (1314) & cat (1376) & damag (923) & event (1331) & enemi (2795)
& deton (377) & attack (620) & local (543)\\
& forc (990)& bda (1045) & bda (908) & saf (1205) & tic (2128) &vest
(294)& close (603) & kill (389) \\
& close (742) & strike (1005) & mm (705) & fire (1071) & contact
(1933) & attack (282) & ie (592) & hospit (385) \\
& friend (737) & kia (849) & pid (667) & aaf (955) & element (1933) &
explos (257) & wia (484) & civilian (357) \\
& isaf (703) & isaf (837) & compund (635) & pax (902) & acm (1635)&
nds (222) & cp (471) & injur (280) \\
& insurg (659) & medevac (777) & ground (613) & contact (818) &
receiv (1480) & kill (195) & isaf (466) & child (239) \\
& track (593) & vehicl (721)& kill (576) & vc (666)& saf (1361) &
khowst (126) & qrf (294) & nation (232) \\
& event (574) & struck (590) & ah (368) & station (653) & arm (1148)
& svbi (112) & checkpoint (263) & lns (216) \\
\hline
\end{tabular*}
\end{sidewaystable}

\section{Computational details}
\label{app2} All calculations have been carried out with the
statistical software \textsf{R} 2.12.0-2.15.1 [\citet{R}] on
\texttt{cluster@WU} [\citet{cluster}]. Topic models were estimated with the
extension package \textbf{topicmodels} 0.0-7 [\citet{Gruen2011}]. Further
packages used were \textbf{slam} 0.1-18 and \textbf{tm} 0.5-4.1. Recursive
partitioning infrastructure was provided by the function \texttt{mob()}
[\citet{zeileis08}] from the package \textbf{party} 0.9-99991. Further
packages used were \textbf{strucchange} 1.4-3. The negative binomial
family model for \texttt{mob} can be found in the package \textbf{mobtools}
0.0-1 [\citet{mobtools}]. It uses \texttt{glm.nb()} in package \textbf{MASS}
7.3-7 [\citet{venables02}].
\end{appendix}

\section*{Acknowledgments}

The authors want to thank Bettina Gr\"un, Torsten Hothorn, Matt Taddy
and Achim Zeileis for discussions, advice and help at various stages of
preparing the manuscript. Additionally, we thank our Editor Susan
Paddock, an anonymous Associate Editor and an anonymous reviewer for
many valuable comments and suggestions which greatly improved the paper.

\begin{supplement}
\sname{Supplement A}\label{suppA}
\stitle{Data, code and plot}
\slink[doi]{10.1214/12-AOAS618SUPPA} 
\sdatatype{.zip}
\sfilename{aoas618\_suppa.zip}
\sdescription{A bundle containing the data sets, the code files and a
high-resolution version of Figure \ref{figall}.}
\end{supplement}

\begin{supplement}
\sname{Supplement B}\label{suppB}
\stitle{Model validation}
\slink[doi]{10.1214/12-AOAS618SUPPB} 
\sdatatype{.pdf}
\sfilename{aoas618\_suppb.pdf}
\sdescription{A detailed description of our validation steps and their
results.}
\end{supplement}


\printaddresses

\end{document}